# Thermal radiation of Er doped dielectric crystals: Probing the range of applicability of the Kirchhoff's law

Ekembu K. Tanyi [1], Brandi T. Burton [1], Evgenii E. Narimanov [2], and M. A. Noginov [1*].

[1] *Center for Materials Research, Norfolk State University, Norfolk, VA 23504*
[2] *Birck Nanotechnology Center, Department of Electrical and Computer Engineering, Purdue University, West Lafayette, IN 47907, USA*
*[*]mnoginov@nsu.edu*

**Abstract:** The Kirchhoff's law of thermal radiation, relating emissivity $\varepsilon$ and absorptance $\alpha$, has been originally formulated for opaque bodies in thermodynamic equilibrium with the environment. However, in many systems of practical importance, both assumptions are often not satisfied. In this work, we revisit the century-old law and examine the limits of its applicability in an example of Er:YAG and Er:YLF dielectric crystals – potential radiation converters for thermophotovoltaic applications. In our experiments, the (80 at.%)Er:YAG crystal was opaque between 1.45 μm and 1.64 μm. In this spectral range, its absorbtance $\alpha(\lambda)$ is spectrally flat and differentiates from unity only by a small amount of reflection. The shape of the emissivity spectrum $\varepsilon(\lambda)$ closely matches that of absorptance $\alpha(\lambda)$, suggesting that the Kirchhoff's law can adequately describe thermal radiation of opaque bodies, even if the requirement of thermodynamic equilibrium is not satisfied. The (20 at.%)Er:YLF crystal had smaller size, lower concentration of Er ions, and it was not opaque. Nevertheless, its spectrum of emissivity $\varepsilon(\lambda)$ had almost the same shape (between 1.45 μm and 1.62 μm) as the spectrum of absorptance $\alpha(\lambda)$ derived from the transmission measurements. This observation is in line with our prediction that the spectra of emissivity and absorptance should have identical shapes in optically thin slabs. We, thus, show that the Kirchhoff's law of thermal radiation can be extended (with caution) even to not-opaque bodies away from the thermodynamic equilibrium.

# 1. INTRODUCTION

## 1.1 Photovoltaics and thermophotovoltaics

In recent years, the development of technologies aimed at harvesting renewable energy, in particular the solar energy [1, 2], has become critically important. The solar energy, most commonly, is absorbed and converted to electricity in a p-n junction semiconductor device known as photovoltaic solar cell [3]. However, photovoltaic cells face an inherent dilemma. Thus, if the bandgap of the semiconductor is large, then a significant portion of the (long-wavelength) sunlight radiation cannot be absorbed, reducing the efficiency of the energy production (Fig. 1a). On the other hand, in a small-bandgap semiconductor, large fraction of the absorbed photons' energy is converted to heat that lowers the efficiency of the light-electricity conversion as well (Fig. 1b). In the seminal publication [4], Shockley and Queisser have shown that for the sunlight spectrum approximated by the blackbody radiation at T=6000 K, the ultimate efficiency of the light-to-electrical energy conversion by a single p-n junction device is equal to only 44 %, at the optimal semiconductor bandgap equal to 1.1 eV (very close to that of Si [5]) and the solar cell temperature equal to 0 K. This value is reduced down to ~31% at the bandgap equal to 1.36 eV, if the realistic temperature, impedance matching, geometrical factors, etc. are taken into account [4].

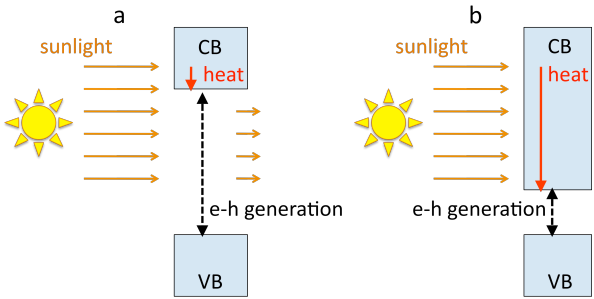

FIG. 1. Schematics of generation of electron-hole (e-h) pairs, showing low efficiency of light absorption in a large-gap semiconductor (a) and large heat production in a low-gap semiconductor (b). (VB and CB stand for valence band and conduction band, respectively.)

The Shockley-Queisser limit can be overcome in tandem schemes, in which solar cells with smaller band gaps are placed behind the solar cells with larger band gaps [6-10]. However, technological complexity and high cost prohibit widespread commercial use of such devices.

An ideal photovoltaic cell should harvest solar energy in the whole spectrum and utilize most of it to create electron-hole (e-h) pairs. Such functionality can be achieved in a low-gap semiconductor combined with so-called thermophotovoltaic converter [11-18]. The role of the converter is to absorb the solar energy over a broad spectrum, become hot, and reemit the absorbed energy in form of thermal radiation in a narrow spectral band matching the bottom of the semiconductor's conduction band, Fig. 2. A significant effort has been made to design sub-wavelength patterned surfaces (metasurfaces) [19-22] or bulk engineered composite materials (metamaterials) [23-31] able of absorbing solar radiation on one side of the converter and reemitting in a narrow infrared band on the other side.

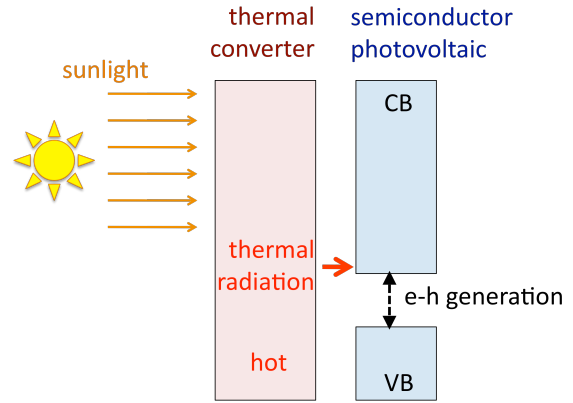

Fig. 2. Schematics of a thermophotovoltaic device.

Alternatively, thermal converter of a thermophotovoltaic device can be based on an inorganic crystal doped with rare-earth [32-36] and transition metal ions [34,37]. Refractory oxide crystals of the garnet family doped with $Er^{3+}$ ions are appealing for thermophotovoltaic applications [32,35] because they have strong absorption in the visible part of the spectrum, strong emission at ~1.55 μm (Fig. 3), and high melting temperature (1965°C for $Er:Y_3Al_5O_{12}$) [38-40]. Even higher absorption efficiency can be achieved in $Er^{3+}$ doped crystals, such as yttrium scandium gallium garnet ($Y_3Sc_2Ga_3O_{12}$ or YSGG), co-doped by $Cr^{3+}$ ions [39,41]. Other rare-earth ions of potential importance to thermophotovoltaics have strong absorption and emission bands at ~1.8 μm ($Tm^{3+}$), ~2.0 μm ($Ho^{3+}$) and ~3.4 μm ($Dy^{3+}$) [38,39,42]. The focus of this paper is on the crystals doped with $Er^{3+}$ ions.

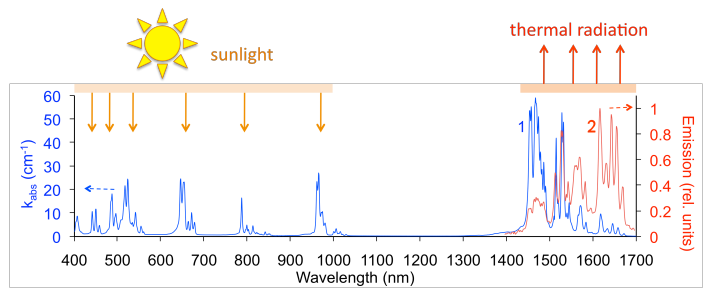

Fig. 3. Room temperature absorption (blue trace 1) and emission (red trace 2) spectra of (80 at.%) Er:YAG; optical path $l$=1.22 mm.

## 1.2 Rare-earth based thermophotovoltaic converters: Kirchhoff's law of emission and absorption

Of particular importance are the absorption and emission spectra of a thermal converter. These functions are related to each other by the Kirchhoff's law [43-45] stating that for *opaque* bodies in *thermodynamic equilibrium*, the radiant emissivity $\varepsilon$ (defined as the ratio of emitted radiant power $P$ to that from an ideal blackbody $P_{BB}$ at the same temperature) is equal to radiant absorptance $\alpha$ (defined as the ratio of absorbed radiant power to incident radiant power) [46], $\varepsilon=\alpha$. The latter quantity is equal to $\alpha=1-\rho-\tau$, where $\rho$ is the radiant reflectance (the ratio of absorbed radiant power to incident radiant power) and $\tau$ is the radiant transmittance (the ratio of transmitted radiant power to incident radiant power); in opaque bodies, $\tau=0$. This law can be formulated for each particular temperature $T$, frequency $\omega$, polarization $\varphi$, and polar coordinates $(\theta,\phi)$ defining the angle of incidence [47]:

$$\varepsilon(\omega,\varphi,\theta,\phi) = \alpha(\omega,\varphi,\theta,\phi), \tag{1a}$$

$$\varepsilon(\omega,\varphi,\theta,\phi) = \frac{P(\omega,\varphi,\theta,\phi,T)}{P_{BB}(\omega,\varphi,\theta,\phi,T)}, \tag{1b}$$

$$\alpha(\lambda,\varphi,\theta,\phi) = 1-\rho(\omega,\varphi,\theta,\phi)-\tau(\omega,\varphi,\theta,\phi) = 1-\sigma(\omega,\varphi,\theta,\phi), \tag{1c}$$

where $\sigma$ is the scattering, defined to take into account all reflected and transmitted light.

The comparative studies of absorption and thermal radiation spectra of three-valent rare-earth ions doped into a variety of oxide hosts have been undertaken in the first half of the 20th century [48-50]. Thus, Wood [49], using the Bausch and Lomb one-prism spectrograph and suitable color filters, has shown that in thin quartz rods doped with $Nd^{3+}$ ions, the absorption bands (approximately) correspond to the thermal emission bands [49,50]. Qualitatively similar behavior has been observed in pulverized and fused erbium oxide [48,50]. Given the technology of the time, only the maxima of the spectral lines could be analyzed more or less accurately, but neither the line-shapes nor the line intensities. Correspondingly, the precise spectra of radiant emissivity $\varepsilon$ could not be obtained and compared with the spectra of radiant absorptance $\alpha$.

In this work, we study the spectra of absorptance $\alpha$, thermal radiant power $P$, and emissivity $\varepsilon$ of dielectric crystals doped with $Er^{3+}$ ions (potential thermophotovoltaic converters), their agreement with the Kirchhoff's law of thermal radiation and the limits of the law's applicability.

## 2. RESULTS

### 2.1 Model considerations

Let us assume that a crystal doped with rare-earth ions in high concentration has its maximal absorption coefficient in the visible-to-mid-infrared spectral range equal to $k_{abs}^{\max}=100$ cm$^{-1}$ and minimal absorption coefficient equal to $k_{abs}^{\min}=0.01$ cm$^{-1}$ – typical values for rare-earth doped laser crystals [40,51]. The absorption coefficient $k_{abs}^{\max}=100$ cm$^{-1}$ is too small to cause any significant change of the real part of the refractive index $\eta$, which, in the first approximation, can be assumed to be dispersion-less.

It is instructive to examine thermal emission and absorption occurring at three different length-scales.

(1) When the size of the radiating body is large, $l >> (k_{abs}^{\min})^{-1}=1$ m, then the body is opaque, $\tau \to 0$. Far from its edges, the body should have an appearance of an anthracite coal – black and moderately shiny. Following the Kirchhoff's law (and easing the requirement of thermodynamic equilibrium), the emissive power $P$ is predicted to be that of the blackbody $P_{BB}$ multiplied by the (nearly spectrally independent) factor $(1-\rho)$,

$$P = P_{BB}(1-\rho). \tag{2}$$

(2) At the intermediate length-scale, $(k_{abs}^{\min})^{-1} > l \approx 1$ cm $> (k_{abs}^{\max})^{-1}$, the body is nearly opaque in the spectral range of strong absorption, $l > (k_{abs}^{\max})^{-1}$, and nearly translucent in the spectral range of small absorption, $l < (k_{abs}^{\min})^{-1}$. This length-scale appears to be the most relevant to thermophotovoltaic devices based on rare-earth thermal converters. However, since the body is not completely opaque, it is questionable whether the Kirchhoff's law, as formulated above, can be directly applied to this case as well as case 3 below. This gap of knowledge and lack of available working model motivated the present study of thermal radiation of not (completely) opaque bodies.

(3) When the size of the body is small, $l << (k_{abs}^{\max})^{-1}=100$ μm, $\alpha \to 0$ and the body is translucent. This regime is easier to realize. Ignoring wave phenomena and considering light to be a stream of photons, the absorptance $\alpha(\omega)$ is predicted to be equal to

$$\alpha(\omega) = 1-\rho(\omega)-\tau(\omega) \approx k_{abs}(\omega)l \tag{3}$$

(see Section 1 of Supplementary Materials). Presuming that the Kirchhoff's law can be extended to not opaque media and combining Eqs. 1a and 3, one obtains

$$\varepsilon(\omega) = \alpha(\omega) \approx k_{abs}(\omega)l. \tag{4}$$

Note that the latter equation can be derived (see Supplementary Materials (Section 1)) using a simple spectroscopic model, which assumes that the spectra $k_{abs}(\omega)$, $P(\omega)$ and $\varepsilon(\omega)$ are comprised of the four lines originating from the transitions between two Stark components of the ground state ($|g_1\rangle$ and $|g_2\rangle$) and two Stark components of the excited state ($|e_1\rangle$ and $|e_2\rangle$), Fig. 4a. If, for simplicity, all four transitions have the same spectral widths and cross sections, then the strengths of the absorption lines $k_{abs}(\omega)$ are determined by Boltzmann populations of the ground state components $|g_1\rangle$ and $|g_2\rangle$ (Fig. 4b and Eq. A5). At the same time, the line intensities in the *very different* spectrum of emitted radiant power $P(\omega)$ are proportional to the Boltzmann populations of the excited state components $|e_1\rangle$ and $|e_2\rangle$ (Fig. 4c and Eq. A9).

However, according to the Kirchhoff's law (Eq. 1), the spectrum of absorptance $\alpha(\omega)$ should be compared not to the spectrum of radiant power $P(\omega)$ but rather to the spectrum of emissivity $\varepsilon(\omega) \equiv P(\omega)/P_{BB}(\omega)$. The latter (Fig. 4d and Eq. S12) was shown to have exactly the same shape as the spectrum $\alpha(\omega)$ (Fig. 4b and Eq. S9), see Supplementary Materials (Section 1).

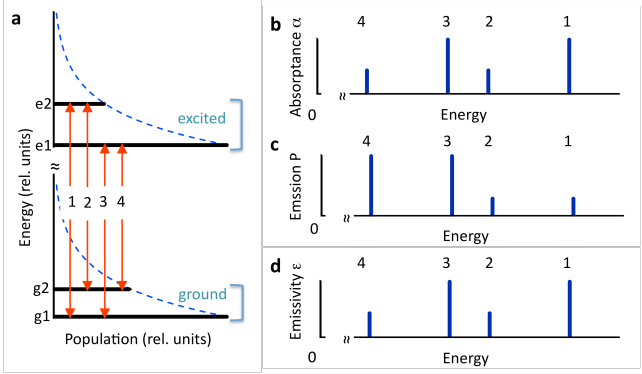

Fig. 4. (a) Schematics of the energy states and Boltzmann distribution of population. Dashed line is the Boltzmann function. The population of the ground state components $|g_1\rangle$ and $|g_2\rangle$ is not in scale with the population of the excited state components $|e_1\rangle$ and $|e_2\rangle$. (b) Relative intensities of the absorptance spectral lines ($\alpha \propto k_{abs}$). (c) Relative intensities of the emission spectral lines (radiant power $P$). (d) Relative intensities of emissivity spectral lines $\varepsilon$. (Same as for absorptance lines in Fig. 4b).

The above mentioned derivation was based on the assumption that $k_{abs}l<<1$ and no photons emitted deep in the volume of the medium are getting re-absorbed and re-emitted as they travel to the surface. In optically thick bodies, in the spectral ranges characterized by $k_{abs}l\geq1$, the photons emitted in the volume of the sample are re-absorbed and partly re-emitted by the medium. Therefore, the photons reaching the surface are emitted reasonably close to the surface (within $l\approx k_{abs}^{-1}$). At the same time, in the spectral ranges with $k_{abs}l<<1$, the photons reaching the surface are generated uniformly throughout the whole volume of the body. This effect is expected to modify the radiation spectrum [52] by suppressing the peaks and enhancing the wings of the emission spectral lines. The experimental studies of thermal radiation of dielectric crystals doped with $Er^{3+}$ ions are presented below.

## 2.2 Experimental results

Two samples used in the thermal radiation studies were Er doped $Y_3Al_5O_{12}$, (80 at.%)Er:YAG, and Er doped LiYF$_4$, (20 at.%)Er:YLF [53,54]. The geometry of the samples and the details of the measurements are discussed in Methods.

*Room temperature spectra.* The room temperature absorption and emission spectra of Er:YAG and Er:YLF are depicted in Fig. 5. In line with the arguments of Section 2.1 (see also Fig. 4 and Section 1 of Supplementary Materials), the distribution of the spectral line intensities in the absorption spectra are different from those in the emission spectra.

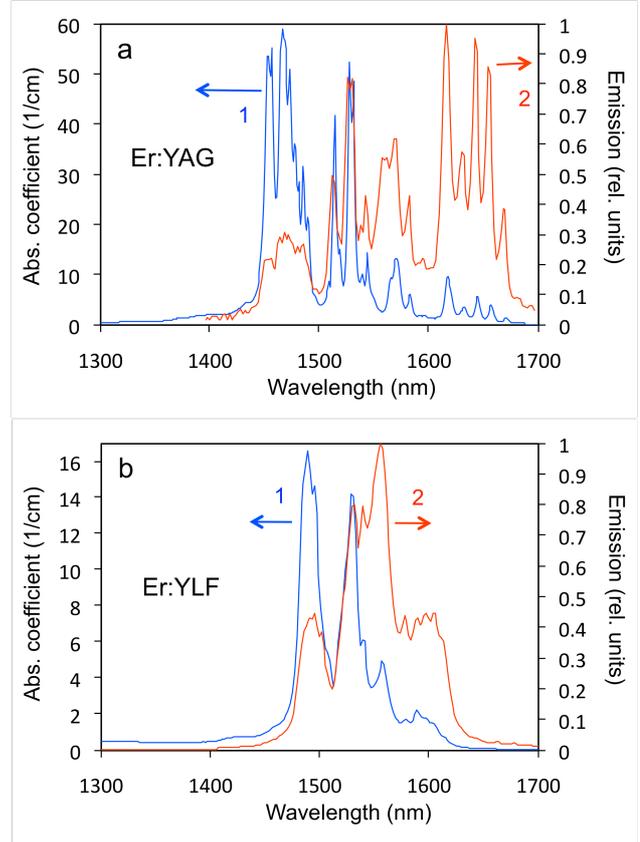

Fig. 5. (a) Room temperature absorption (1) and emission (2) spectra of (80 at.%)Er:YAG (optical path $l$=1.2 mm). (b) Same for (20 at.%)Er:YLF (optical path $l$=3.0 mm).

*High temperature spectra.* The spectra of thermal radiation, transmittance and scattering of the heated Er:YAG and Er:YLF crystals have been acquired and used to derive the spectra of emissivity $\varepsilon(\lambda)$ and absorptance $\alpha(\lambda)$ as described in Methods. The spectra of emitted radiant power $P(\lambda)$ are shown as traces 2 in Figs. 6a and 6b, respectively, and the corresponding emissivity spectra $\varepsilon(\lambda)$ are depicted as traces 1 in Figs. 7a and 7b. (As follows from the discussion of the experimental setups and the normalization procedure in Methods (see also Figs. 8 and 9), only the shapes of the emissivity spectra, but not the absolute values $\varepsilon(\lambda)$ can be analyzed in our experiment.)

The spectrum of absorption coefficients $k_{abs}(\lambda)$ (Fig. 6) as well as the spectrum of absorptance $\alpha(\lambda)$ (Fig. 7) have been acquired as described in Methods and Section 2 of Supplementary Materials.

In the Er:YAG crystal, multiple angular absorptance spectra $\alpha(\lambda,\theta_i,\phi_i,\theta_j,\phi_j)$ calculated (as discussed in Methods and Section 2 of Supplementary Materials) from the scattering spectra $\sigma(\lambda,\theta_i,\phi_i,\theta_j,\phi_j)$, had nearly the same shape, Fig. 7a, trace 3. They were also reasonably close to the absorptance spectrum (Fig. 7a, trace 2) recorded in the transmission setup of Fig. 9c. In Er:YLF, the spectra $\alpha(\lambda,\theta_i,\phi_i,\theta_j,\phi_j)$ obtained from the scattering measurements carried out at different orientations of the sample (Fig. 7b, trace 3) were also reasonably similar to each other. However, none of them resembled the absorptance spectrum measured in the transmission experiments (Fig. 7b, trace 2). (Note that the experimental procedure discussed in Methods allowed one to deduce only the spectral shape of absorptance $\alpha(\lambda)$ but not its magnitude.)

The spectra of Figs. 5,6 and 7 are discussed in detail Section 4 below.

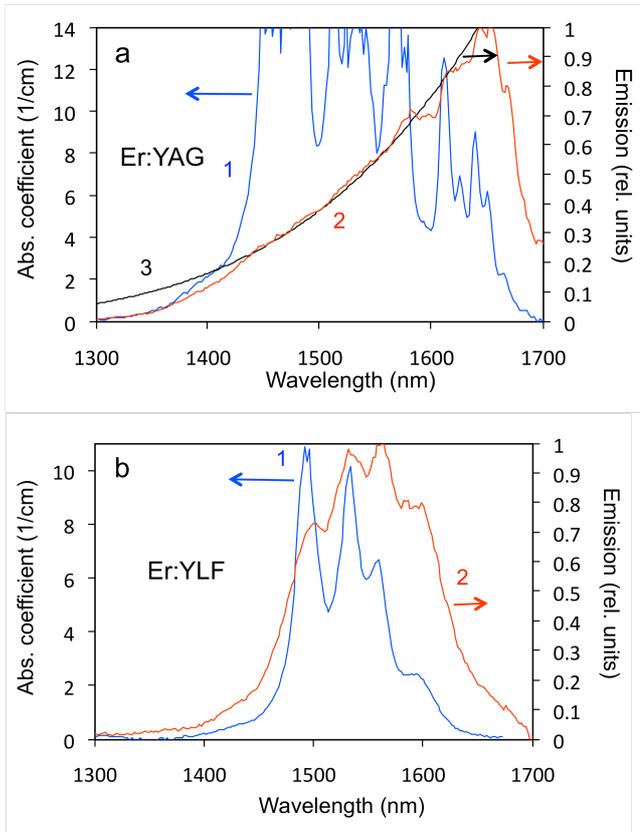

Fig. 6. (a) High temperature (551 K) absorption spectrum (trace 1) and high temperature (583 K) emission spectrum (trace 2) of the (80 at.%)Er:YAG (optical path $l$=3.8 mm). Black body emission spectrum calculated at the sample's temperature 583K (trace 3). Because of the large thickness of the crystal, the absorption coefficients in the peaks of the spectral lines ($\geq 14$ cm$^{-1}$) could not be determined accurately (not shown). (b) High temperature (612 K) absorption spectrum (trace 1) and high temperature (588 K) emission spectrum (trace 2) of the (20 at.%)Er:YLF (optical path $l$=3.0 mm).

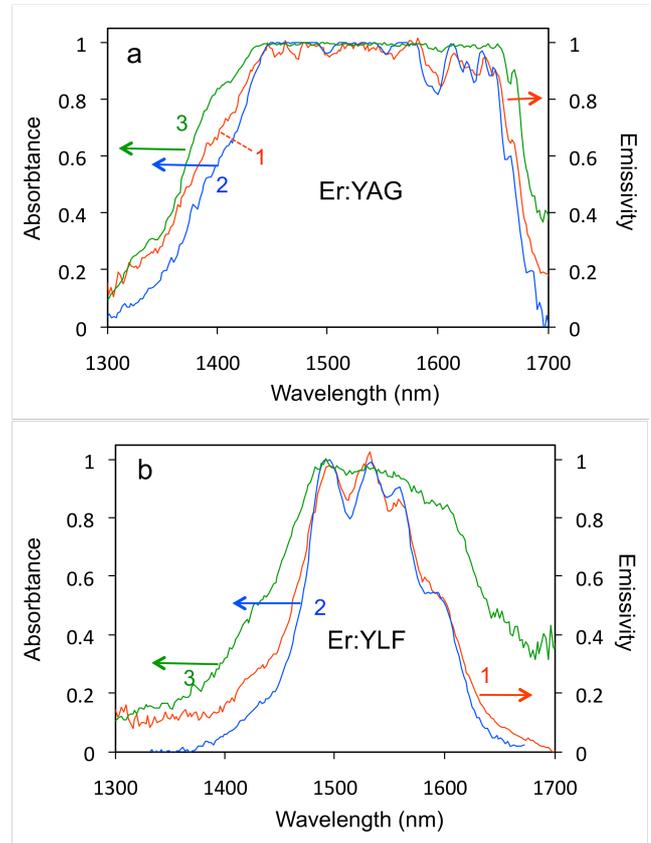

Fig. 7. (a) Emissivity of the Er:YAG crystal (trace 1) and its absorptance measured in the transmission (trace 2) and scattering (trace 3) experiments. Traces 2 and 3 are normalized to unity. Trace 1 was scaled to have the best match with trace 2 between 1.44 μm and 1.65 μm. (b) Same for the Er:YLF crystal.

## 4. DISCUSSION

### 4.1 Trends in emission and absorption

In line with the arguments of Section 1 of Supplementary Materials (lower Stark components of the $Er^{3+}$ ground state $^4I_{15/2}$ and the excited state $^4I_{13/2}$ are populated more than the higher energy components), the stronger absorption lines are expected in the short wavelength range of the $^4I_{15/2} - {}^4I_{13/2}$ spectrum while the strongest emission lines are expected to be found at longer wavelengths. This is clearly the case of the Er:YAG and Er:YLF crystals, whose room temperature absorption and emission spectra are depicted in Figs. 5a and 5b.

With increase of the temperature, high-energy components of the $^4I_{15/2}$ and $^4I_{13/2}$ multifolds become more populated. In result, the center of gravity of the absorption spectrum is expected to shift toward longer wavelengths and that of the emission spectrum is supposed to shift toward shorter wavelengths. This trend is seen in the absorption and thermal radiation spectra of Er:YLF (Fig. 6b) and the absorption spectrum of Er:YAG (Fig. 6a). The behavior of the thermal radiation spectrum of Er:YAG is noteworthy. It does not show much of the blue shift, predicted based on the population arguments above. At the same time, in the broad spectral range, it closely follows the spectrum of the blackbody emission calculated at the same temperature, compare traces 2 and 3 in Fig. 6a. The significance of this behavior is emphasized in Section 4.2 below.

The spectra of absorption (absorption coefficients) and emission (emitted power) in Figs. 5 and 6 are not identical to each other and they should not be identical (see Section 1 of Supplementary Materials). According to the Kirchhoff's law, the quantities, which have to be equal to each other, are the radiant absorptance $\alpha(\lambda)$ and the radiant emissivity $\varepsilon(\lambda)$. The agreement of the experimental absorptance and emissivity spectra is discussed below.

### 4.2 Absorptance and emissivity of Er:YAG

The Er:YAG crystal used in our experiments has higher concentration of $Er^{3+}$ ions and larger physical size than the Er:YLF crystal. It is nearly opaque and its transmittance is very close to zero between ~1.44 µm and ~1.65 µm, Fig. 7a. In the latter spectral range, the absorptance (evaluated in both transmission and scattering experiments) is spectrally flat and differentiates from unity by a small amount of nearly dispersion-less reflection. Remarkably, the shape of the absorptance spectrum $\alpha(\lambda)$ closely matches that of the emissivity $\varepsilon(\lambda)$, which is also nearly flat and featureless. (The latter fact could be anticipated, since the emission spectrum of Er:YAG in Fig. 6a followed that of the blackbody at the same temperature.) In this sense, at small angles of incidence, the crystal behaves *almost* like a blackbody (whose absorptance differentiates from unity by a small reflection coefficient $\rho(\lambda)$). We, thus, have shown that in the spectral range of the sample's opacity, the absorptance $\alpha(\lambda)$ and emissivity $\varepsilon(\lambda)$ are consistent with the predictions of the Kirchhoff's law, even if the heated crystal is not in thermodynamic equilibrium with the room temperature environment. Outside of the opacity range, the agreement between the absorptance and the emissivity is poorer and depends on the geometry in which the absorptance is measured.

### 4.3 Absorptance and emissivity of Er:YLF

The Er:YLF crystal was smaller and had lower concentration of $Er^{3+}$ ions. It was not completely opaque. It had relatively large transmittance in the whole spectral range of interest, and its spectra of emissivity $\varepsilon(\lambda)$ (Fig. 7b, trace 1) and absorptance $\alpha(\lambda)$, retrieved from the transmittance measurements (Fig. 7b, trace 2), featured spectral lines characteristic of $Er^{3+}$ ions. (The latter were partly smeared out due to high temperature.) Remarkably, the latter two traces had nearly similar shapes between 1.45 µm and 1.62 µm. This suggests that the Kirchhoff's law is applicable to not opaque bodies as well. At the same time, the spectrum of absorptance deduced from a typical scattering experiment (Fig. 7b, trace 3) had a completely different shape. One can infer that the photons reaching the detector in the emission, transmission and scattering experiments experience different amounts of external and internal scattering off polished and unpolished surfaces of the sample. This can qualitatively explain the difference between traces 2 and 3 in Fig. 7b.

Note that the emission and absorption spectra in Figs 6a and 6b were recorded at slightly different temperatures. Knowing the Stark splitting of the ground state multiplet $^4I_{15/2}$ and the excited state multiplet $^4I_{13/2}$ at the temperatures measured in the absorption and emission experiments, we estimate that the relative populations of the lowest and the highest Stark levels (and the intensities of the corresponding lines in the absorption and emission spectra) should vary with temperature by ~8% for YAG and ~3% for YLF. These small differences are practically not noticeable in the spectra of absorptance and emissivity in Figs. 7a and 7b. Moreover, the good agreement between the absorptance and the emissivity for both Er:YAG and Er:YLF serves as the evidence of adequately accurate temperature measurements.

### 4.4. Use of $Er^{3+}$ and other rare-earth ions in thermophotovoltaic applications

As dielectric crystals doped with $Er^{3+}$ ions can emit thermal radiation in the relatively narrow near-infrared spectral band, the potential use of these and similar materials in thermophotovoltaic applications [32] deserves serious consideration. We infer that thermophotovoltaic converters based on rare-earth doped materials should have the following desired properties.

(i) The material should be refractory (able to sustain high temperature).
(ii) The material should be opaque in both visible and near-infrared spectral ranges – to efficiently absorb solar energy and to emit thermal radiation. High concentration of $Er^{3+}$ or other rare-earth dopant ions serves this purpose.
(iii) Even at large concentration of rare-earth ions, absorption in the visible spectral range may be too small. Co-doping the crystals with $Cr^{3+}$ ions, which have strong broadband absorption in the visible part of the spectrum, as in *e.g.* yttrium scandium gallium garnet ($Y_3Sc_2Ga_3O_{12}$) doped with Er and Cr [39,41], should improve absorption of solar radiation significantly.
(iv) Thermal radiation band in the infrared range of the spectrum should be narrow. Therefore, the material should have weak crystal field and small Stark splitting of both ground and excited states [39].
(v) Choosing the dopant ions, which have absorption and thermal radiation bands at longer wavelengths, such as $Tm^{3+}$ ($\lambda \approx 1.8$ μm), $Ho^{3+}$ ($\lambda \approx 2.0$ μm) and $Dy^{3+}$ ($\lambda \approx 3.4$ μm) [38], can substantially decrease the operating temperature of the thermophotovoltaic converter and increase its efficiency.

Since not all the desired properties above may be available in one material, engineering of the thermophotovoltaic converter may require thorough optimization.

## 5. SUMMARY

The Kirchhoff's law of thermal radiation predicts that the radiant emissivity $\varepsilon$ (the ratio of emitted radiant power to that from an ideal blackbody at the same temperature) of *opaque* bodies in ther*modynamic equilibrium* with the environment should be equal to absorptance $\alpha$ (the ratio of absorbed power to incident power). At the same time, many radiating bodies of practical importance are not opaque and not in thermodynamic equilibrium. The purpose of this work was to study the limits of applicability of the Kirchhoff's law in an example of Er:YAG and Er:YLF dielectric crystals – potential radiation converters for thermophotovoltais applications.

Using a simple model, we have shown that the spectra of emissivity $\varepsilon$ and absorptance $\alpha$ (at normal incidence) should have identical shapes in optically thin slabs at $k_{abs}l \ll 1$, without requirement of thermodynamic equilibrium with the environment (only thermalization within the ground state multifold and the excited state multifold is needed).

Experimentally, we have found the 3.8 mm thick (80 at.%) Er:YAG crystal to be nearly opaque between 1.44 μm and 1.65 μm. In this spectral range, its absorptance $\alpha(\lambda)$ is spectrally flat and differentiates from unity only by a small amount of reflection. One can say that at small angles of incidence, the crystal behaves almost like a blackbody. The shape of the emissivity spectrum $\varepsilon(\lambda)$ closely matches that of absorptance $\alpha(\lambda)$, confirming that the Kirchhoff's law can adequately describe thermal radiation of opaque bodies, even if the requirement of thermodynamic equilibrium with the environment is not satisfied.

The (20 at.%) Er:YLF crystal had smaller size (3.0 mm thick), lower concentration of Er ions, and it was not opaque. Nevertheless, the spectrum of emissivity in this sample $\varepsilon(\lambda)$ had almost the same shape as the spectrum of absorptance $\alpha(\lambda)$ derived from the transmission measurements. This suggests that the predictions of the Kirchhoff's law can be extended to not opaque bodies, which are not in thermodynamic equilibrium with the environment. At the same time, the spectrum $\alpha(\lambda)$ retrieved from the scattering measurements had completely different shape. Therefore, in not opaque bodies, the spectra $\alpha(\lambda)$ and $\varepsilon(\lambda)$ may coincide or not, depending on the geometry of the experiment as well as scattering-dependent photon residence time in the sample.

Lastly, rare-earth doped dielectric crystals appear to be promising radiation converters for thermophotovoltaic applications, which deserve further studies.

## 6. METHODS

### 6.1 Experimental samples

Two samples used in the thermal radiation studies were Er doped $Y_3Al_5O_{12}$ (Er:YAG) and Er doped $LiYF_4$ (Er:YLF) [53,54]. According to the manufacturer [53], the concentration of $Er^{3+}$ ions in the YAG crystal was equal to 80 at. %. This is consistent with the published spectroscopic data [40,55]. The $Er^{3+}$ concentration in the Er:YLF crystal, determined by comparing the absorption spectrum with the literature sources [56], was equal to ~20 at. %. The shapes and sizes of the crystals are shown in Fig. 8a. The Er:YAG crystal was too thick to accurately measure the absorption coefficients in the maxima of the spectral lines. Therefore a thin, $l=1.2$ mm, plate (with polished parallel sides) has been used in the spectrophotometer and spectrofluorimeter measurements as discussed below.

### 6.2 Room temperature spectral measurements

Room temperature absorption spectra of the Er:YAG and Er:YLF crystals have been collected using Lambda 900 spectrophotometer equipped with the 150 mm integrating sphere from Perkin Elmer. The spectra have been corrected for reflection. The room temperature emission spectra have been recorded using FluoroLog 3 Modular Spectrofluorometer (from HORIBA Yobin Ivon).

### 6.3 High temperature spectral measurements

*Experimental setups.* In the high-temperature spectral measurements, the crystals were mounted on a flat top of the solder iron rod (6 mm in diameter) wrapped with aluminum foil to reduce the thermal radiation, Fig. 8b. The rod was heated with the modified 40 W solder iron from Weller. The supplied voltage was controlled by the auto-transformer. A chromel-alumel thermocouple (K type, from OMEGA Engineering) was pressed against the top surface of the sample. In order to simulate the blackbody emitter, black soot was deposited onto the solder iron tip [57,58].

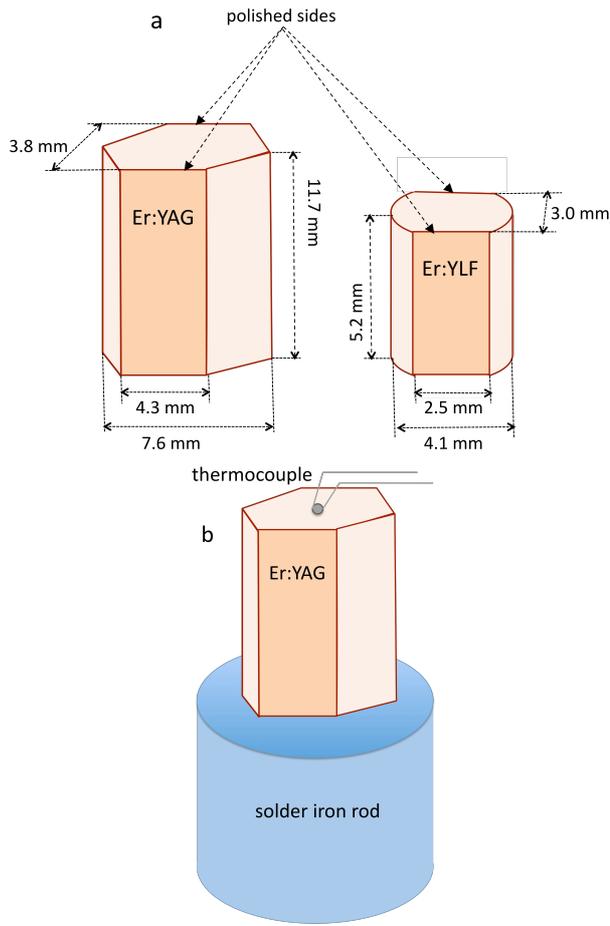

Fig. 8. (a) Geometry of the Er$^{3+}$ doped crystals used in the thermal radiation studies. (b) Crystal mounted on top of the solder iron rod (heater).

Three types of high temperature spectral measurements have been performed: emission, transmittance, and scattering. The corresponding experimental setups are shown in Fig. 9.

The optical spectra have been recorded using the MS257 monochromator (equipped with the 1000 grooves/mm grating) and Merlin lock-in amplifier (from Newport/Oriel). The wavelength of the monochromator was calibrated using 632.8 nm line of the He-Ne laser, and the spectral response of the apparatus was calibrated using the spectrum of the black body (solder iron tip coated with black soot) recorded at the known temperature. The signal was detected with the peltier-cooled InGaAs photodiode (Electrical Optical Systems Inc.), whose spectral sensitivity range spanned from 0.95 μm to 1.75 μm. The 50W halogen lamp served as the light source in the transmission and scattering measurements.

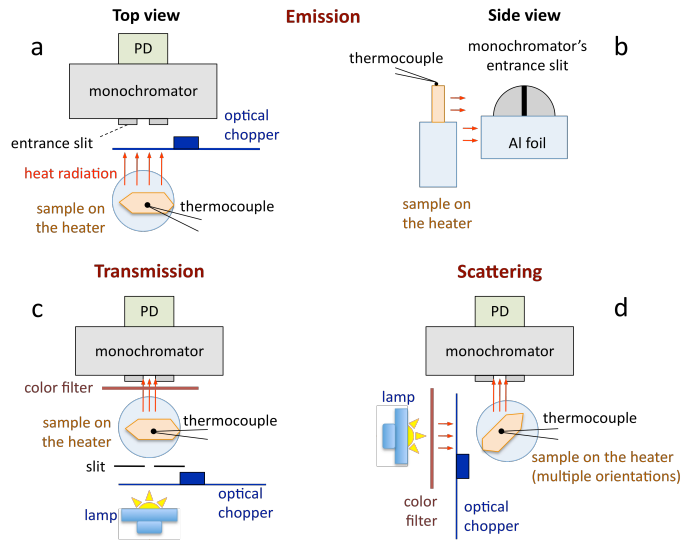

Fig. 9. (a) Emission measurement setup (top view). (b) Sample on the heater stage in front of the entrance slit of the monochromator, which is partly covered with the aluminum foil in the emission measurements (side view). (c) Transmission measurement setup (top view). (d) Scattering measurement setup (top view).

*Emitted radiant power and emissivity.* In the emission spectral measurements, the distance between the sample and the entrance slit of the monochromator was equal to ≈1 cm. The optical chopper (providing reference signal to the lock-in amplifier) was installed between the sample and the detector, Fig. 9a. In order to minimize the amount of thermal radiation of the solder iron rod (heater) getting to the entrance slit of the monochromator, the bottom half of the slit was covered with the aluminum foil, Fig. 9b. In the two consecutive measurements, we have recorded the spectrum of thermal radiation of (*i*) the sample on the heater stage and (*ii*) the heater stage without the sample. To obtain the thermal radiation spectrum of the sample without contribution from the heated environment, the (much weaker) latter signal was subtracted from the former one. When the resultant spectrum was divided by the response function of the apparatus, the spectrally corrected spectrum of the emitted radiant power $P$ has been generated. To obtain the spectrum of radiant emissivity $\varepsilon$, the spectrum of emitted radiant power of the sample $P$ was divided by that of the blackbody $P_{BB}$ calculated at the same temperature.

Note that the thermal radiation spectra of Er doped samples and the black body were recorded in different geometries. (In the black body measurements, the solder iron tip covered with the black soot was oriented horizontally, the monochromator's slit was not covered with aluminum foil, the size of the soldering iron rod was different from that of the sample, and the distance between the soldering iron rod and the slit of the monochromator was slightly different from that in the Er-doped sample measurements.) Therefore, only the spectral shapes of the emissivity spectra $\varepsilon(\lambda)$, but not the absolute emissivity values can be analyzed.

*Transmission and absorption.* The transmission spectral measurements were carried out using the setup depicted in Fig. 7c. The optical chopper was installed between the lamp and the crystal. Therefore, thermal radiation of the sample did not contribute to the output signal of the lock-in amplifier. The external slit installed several millimeters away from the sample ensured that no photons emitted by the lamp reached the entrance slit of the monochromator bypassing the crystal. The long-pass color filter blocked the (second order of diffraction) visible light, which otherwise would mix with the (first order of diffraction) infrared light to be studied. In a typical experiment, two spectra have been recorded, with and without the sample. The presence of the crystal slightly reshaped the light beam (possibly due to unintentional curving of the nominally flat parallel polished crystal surfaces, which produced the lensing effect). Therefore, after dividing the former spectrum by the latter, the resultant transmittance spectrum was further multiplied by a correction factor, bringing the transmission in the spectral ranges, which did not have any absorption, to 100%.

*Scattering.* In the scattering measurements, the crystal was installed in front of the entrance slit of the monochromator and illuminated from the side, Fig. 7d. The optical chopper was installed between the lamp and the crystal. Therefore, the thermal radiation of the sample did not contribute to the scattering (or the reflection) spectrum. The long-pass color filter was used to block the visible light. In multiple measurements, the crystal was positioned at various angles relative to the lamp $(\theta_i,\phi_i)$ and the monochromator's slit $(\theta_j,\phi_j)$ and was turned to them with its polished or unpolished sides. Correspondingly, in different particular measurements, the specular reflected light and the diffused light reaching the detector were mixed in different proportions and the photons experienced larger or smaller number of scattering events before they exited the samples.

In the control measurement, the crystal was replaced with the broadband white diffused reflector (from Lab Sphere). Experimentally, the two spectra were collected, one with the crystal and one with the diffused reflector. Then the former was divided by the latter, and the resultant scattering spectrum was scaled to reach unity in the spectral ranges where the crystals did not have any absorption (as it is explained above and in Section 2 of Supplementary Materials). This angular-dependent scattering spectrum $\sigma(\lambda,\theta_i,\phi_i,\theta_j,\phi_j)$ was used to obtain the spectrum of the angular sample's absorptance for given directions of illumination and data collection, $\alpha(\lambda,\theta_i,\phi_i,\theta_j,\phi_j) = 1 - \sigma(\lambda,\theta_i,\phi_i,\theta_j,\phi_j)$. (This procedure allowed one to deduce only the spectral shape of absorptance but not its magnitude.)

## Acknowledgments

**General:** The authors cordially thank E. V. Zharikov, H. P. Jenssen and A. Cassanho for providing experimental samples.

**Funding:** The work was partly supported by the NSF PREM grant DMR 1205457, ARO grant W911NF-14-1-0639 and NSF IGERT grant DGE-0966188, NSF MRSEC grant DMR 1120923, and Gordon and Betty Moore Foundation.

## Author contributions

**E.K.T.** conducted most of the experiments, played a key role in the data analysis and participated in writing the manuscript. **B.T.B.** participated in the experiments and data analysis. **E.E.N.** played a key role in providing theoretical considerations and insights. **M.A.N.** designed this study; participated in the experiments, data analysis and writing the manuscript.

## Competing Interests

To the best of our knowledge, the authors have no conflicts of interest.

# SUPPLEMENTARY MATERIALS

## S1. ABSORPTANCE AND EMISSIVITY

### S1.1 Reflectance, transmittance, and absorptance

Let us consider the simplest situation when light falls, at normal incidence, onto a slab of material with two parallel ideally polished walls. Ignoring wave phenomena and considering light to be a stream of photons, the reflectance $\rho$ and transmittance $\tau$ are described by the following formula,

$$\rho = r + \frac{(1-r)^2 r \exp(-2k_{abs}l)}{1 - r^2 \exp(-2k_{abs}l)}, \quad (S1)$$

$$\tau = \frac{(1-r)^2 \exp(-k_{abs}l)}{1 - r^2 \exp(-2k_{abs}l)}, \quad (S2)$$

where $\rho$ is the reflectance at the medium/air interface. It is easy to show that at $k_{abs}l \ll 1$,

$$\alpha(\omega) = 1 - \rho(\omega) - \tau(\omega) \approx k_{abs}(\omega)l. \quad (3)$$

The question arises: what are the expected spectra of thermal radiation power $P(\omega)$ and thermal emissivity $\varepsilon(\omega)$ if the Kirchhoff's laws can be extended to not opaque media? In this case, combining Eqs. 1a and 3, one obtains

$$\varepsilon(\omega) = \alpha(\omega) = k_{abs}(\omega)l \quad (4).$$

Alternatively, this equation can be derived using simple spectroscopic considerations as described below.

### S1.2 Absorptance of translucent samples

For simplicity, let us assume that the spectrum $k_{abs}(\omega)$ is comprised of the four lines originating from the transitions between two Stark components of the ground state ($|g_1\rangle$ and $|g_2\rangle$) and two Stark components of the (metastable) excited state ($|e_1\rangle$ and $|e_2\rangle$), Fig. 4a. Correspondingly,

$$\begin{aligned}\alpha(\omega) &\propto k_{abs}(\omega) \\ &= \sigma_1(\Omega_1,\omega)m_{g_1} + \sigma_2(\Omega_2,\omega)m_{g_2}, \\ &+ \sigma_3(\Omega_3,\omega)m_{g_1} + \sigma_4(\Omega_4,\omega)m_{g_2}\end{aligned} \quad (S3)$$

where $\sigma_i(\Omega_i,\omega)$ are the absorption cross section spectra of the four transitions with the maxima at $\Omega_i$ (Fig. 4a), and $m_{g_1}$ and $m_{g_2}$ are the population number densities of the ground state components $|g_1\rangle$ and $|g_2\rangle$.

Let us also assume, for simplicity, that the four transitions have identical line-shapes and cross sections ($\sigma = \sigma_{1(\omega=\Omega_1)} = \sigma_{2(\omega=\Omega_2)} = \sigma_{3(\omega=\Omega_3)} = \sigma_{4(\omega=\Omega_4)}$). In this case, the intensities of the four absorption lines are solely determined by the populations $m_{g_1}$ and $m_{g_2}$, which are governed by the Boltzmann statistics [59,60]

$$\frac{m_i}{m_j} = \exp\left(-\frac{\Delta E_{ij}}{k_B T}\right). \quad (S4)$$

Here $\Delta E_{ij}$ is the energy difference between states $|i\rangle$ and $|j\rangle$, $T$ is the temperature, $k_B$ is the Boltzmann constant, and no degeneracy is assumed. Substituting Eq. S4 to Eq. S3, one obtains

$$\begin{aligned}\alpha(\omega) &\propto k_{abs}(\omega) \\ &\propto \sigma(\Omega_1,\omega) + \sigma(\Omega_2,\omega)\exp\left(-\frac{\Delta E_{g_1 \to g_2}}{k_B T}\right) \\ &+ \sigma(\Omega_3,\omega) + \sigma(\Omega_4,\omega)\exp\left(-\frac{\Delta E_{g_1 \to g_2}}{k_B T}\right)\end{aligned} \quad (S5)$$

The corresponding strengths of the absorption lines are schematically shown in Fig. 4b. In accord with Eq. S5 and the arguments above, the absorption transitions originating from the lower state $|g_1\rangle$ (lines 1 and 3) have larger strengths than those originating from the higher state $|g_2\rangle$ (lines 2 and 4). Therefore, if the Kirchhoff's law is applicable to not opaque bodies ($k_{abs}l \ll 1$), then the spectrum of radiative emissivity $\varepsilon(\omega)$ should resemble that of Fig. 4b.

### S1.3 Emitted radiant power and emissivity of translucent samples

Returning to the arguments of thermal radiation, light emission originates from atomic and molecular transitions between the higher energy states and the lower energy states. The only difference between thermal radiation and other types of emission, e.g. luminescence or spontaneous emission, is the method of excitation of the higher energy states – in the case of thermal radiation, it occurs as the result of thermal motion of atoms or molecules [61].

Treating thermal radiation as a thermally excited spontaneous emission, one would expect the spectrum of radiant power (emitted in volume $V$ and frequency interval $d\omega$) to be given by

$$P(\omega)/V = [A_1 g(\Omega_1,\omega)m_{e_2} + A_2 g(\Omega_2,\omega)m_{e_2} \\ + A_3 g(\Omega_3,\omega)m_{e_1} + A_4 g(\Omega_4,\omega)m_{e_1}]\hbar\omega d\omega, \quad (S6)$$

where $A_i$ are the spontaneous emission rates of the transitions depicted in Fig. 4a, $g(\Omega_i,\omega)$ are the transitions' line-shapes, $\Omega_i$ are the transitions' central frequencies, $m_{e_1}$ and $m_{e_2}$ are the population number densities of the excited state components $|e_1\rangle$ and $|e_2\rangle$, and $\hbar$ is the Planck's

constant. Since the spontaneous emission rate is related to the emission cross section as [62]

$$A_i g(\Omega_i, \omega) = \sigma_i(\Omega_i, \omega) \frac{\omega^2 \eta^2}{\pi^2 c^2}, \quad (S7)$$

Eq. (S7) can be re-written as

$$P(\omega)/V = [\sigma_1(\Omega_1, \omega) m_{e_2} + \sigma_2(\Omega_2, \omega) m_{e_2} + \sigma_3(\Omega_3, \omega) m_{e_1} + \sigma_4(\Omega_4, \omega) m_{e_1}] \frac{\hbar \omega^3 \eta^2}{\pi^2 c^2} d\omega \quad (S8)$$

(where $\eta$ is the refractive index). If all emission spectral lines have identical shapes and cross sections, then the intensities of the emission lines will be determined by the excited state concentrations $m_{e_1}$ and $m_{e_2}$ (whose ratio is different from that of $m_{g_1}$ and $m_{g_2}$) and $\omega^3$. Whether the excited state is in thermal equilibrium with the ground state (as in the case of thermal radiation) or not (as in the case of optically pumped spontaneous emission), the ratio of the concentrations $m_{e_1}$ and $m_{e_2}$ is determined by Eq. (S4). Therefore,

$$\frac{P(\omega)}{V} = m_{e_1} [\sigma(\Omega_1, \omega) \exp\left(-\frac{\Delta E_{e_1 \to e_2}}{k_B T}\right) + \sigma(\Omega_2, \omega) \exp\left(-\frac{\Delta E_{e_1 \to e_2}}{k_B T}\right) + \sigma(\Omega_3, \omega) + \sigma(\Omega_4, \omega)] \frac{\hbar \omega^3 \eta^2}{\pi^2 c^2} d\omega \quad (S9)$$

Correspondingly, the ratios of the spectral line intensities in the emission spectra (Fig. 4c) are distinctly different from those in the absorption spectra (Fig. 4b). In Fig. 4c, it was assumed that $(\Omega_1 - \Omega_4) \ll \Omega_1, \Omega_4$. Therefore, the $\omega^3$ dependence as well as the frequency dependence of the refractive index $\eta$ could be neglected. (The agreement between the absorption and the emission spectra did not improve if the relatively weak dependences above were taken into account.)

However, according to the Kirchhoff's law (Eq. 1), the spectrum of absorptance $\alpha(\omega)$ should be compared not to the spectrum of radiant power $P(\omega)$ but rather to the spectrum of emissivity $\varepsilon(\omega) \equiv P(\omega)/P_{BB}(\omega)$. As the spectral radiance of a blackbody (radiant energy per unit time, unit projected area, unit solid angle and frequency interval $d\omega$) is given by [62]

$$P_{BB}(\omega) = \frac{\hbar \omega^3}{4\pi^3 c^2} \frac{1}{\exp(\hbar\omega/k_B T) - 1}, \quad (S10)$$

the emissivity is proportional to

$$\varepsilon(\omega) \propto [\sigma(\Omega_1, \omega) \exp\left(-\frac{\Delta E_{e_1 \to e_2}}{k_B T}\right) + \sigma(\Omega_2, \omega) \exp\left(-\frac{\Delta E_{e_1 \to e_2}}{k_B T}\right) + \sigma(\Omega_3, \omega) + \sigma(\Omega_4, \omega)] \eta^2 [\exp(\hbar\omega/k_B T) - 1] \quad (S11)$$

At $\hbar \Omega_i \gg k_B T$, small widths of the spectral lines $\delta \omega_i < \Omega_i$, and spectrally independent refractive index $\eta$,

$$\varepsilon(\omega) \propto$$

$$\propto \sigma(\Omega_1, \omega) \exp\left(\frac{E_{\Omega_1}}{k_B T} - \frac{\Delta E_{e_1 \to e_2}}{k_B T}\right) + \sigma(\Omega_3, \omega) \exp\left(\frac{E_{\Omega_3}}{k_B T}\right)$$

$$+ \sigma(\Omega_2, \omega) \exp\left(\frac{E_{\Omega_2}}{k_B T} - \frac{\Delta E_{e_1 \to e_2}}{k_B T}\right) + \sigma(\Omega_4, \omega) \exp\left(\frac{E_{\Omega_4}}{k_B T}\right)$$

$$\propto \sigma(\Omega_1, \omega) + \sigma(\Omega_2, \omega) \exp\left(-\frac{\Delta E_{g_1 \to g_2}}{k_B T}\right) + \sigma(\Omega_3, \omega)$$

$$+ \sigma(\Omega_4, \omega) \exp\left(-\frac{\Delta E_{g_1 \to g_2}}{k_B T}\right) \quad (S12)$$

(since $E_{\Omega_2} - \Delta E_{e_1 \to e_2} = E_{\Omega_4}$, $E_{\Omega_4} + \Delta E_{g_1 \to g_2} = E_{\Omega_3}$, and $E_{\Omega_1} - \Delta E_{e_1 \to e_2} = E_{\Omega_3}$, Fig. 4a), which is identical to the spectral dependence of the absorption coefficient $k_{abs}(\omega)$ and absorptance $\alpha(\omega)$ (compare Eq. S5 and Figs. 4b and 4d).

Therefore, in a simple example, we have demonstrated that the Kirchhoff's law of thermal radiation $\varepsilon(\omega) \equiv P(\omega)/P_{BB}(\omega) \propto \alpha(\omega)$ can be applied to optically thin bodies, $k_{abs}l \ll 1$. Although, by no means, does this derivation serves as a rigorous proof applicable to all not opaque bodies, it suggests that the Kirchhoff's law can likely be applied to a variety of bodies with different degrees of opaqueness, such as doped crystals and glasses.

## S2. ABSORPTANCE OF A SLAB AT NORMAL INCIDENCE ANGLE

As follows from Eqs. 1c, S1 and S2, at normal incidence of light onto a slab with parallel polished walls, the absorption coefficients $k_{abs}(\lambda)$ and the absorptance $\alpha(\lambda)$ can be determined precisely only if both transmittance $\tau(\lambda)$ and reflectance $\rho(\lambda)$ spectra are known. In practice, transmittance spectra are routinely measured in standard (*e.g.* spectrophotometer) experiments, while the samples' reflectance is available much more rarely. Here we show that for the absorption strengths and physical sizes characteristic of our Er doped crystals, the *approximate* method described below, based on knowledge of the transmittance spectra $\tau(\lambda)$ and the refractive index $\eta$, can be used to determine $k_{abs}$ and $\alpha$ with high degree of accuracy.

In the example below, let us assume that the *exact* absorption spectrum $k_{abs}^e(\lambda)$ (given by *e.g.* trace 1 in Fig. S1a) and the refractive index $\eta$=1.806 (which is assumed to be constant in the whole range of interest, 1.3 μm to 1.7 μm) are known and that the absorption lines are too weak to affect the sample's normal incidence reflectance $r$=0.0825 at the single air/medium interface. (For the purpose of this discussion, $k_{abs}^e(\lambda)$ can be any plausibly looking spectrum characteristic of the samples studied.) Then, using Eqs. S1, S2 and 1c, we can model the *exact* spectra $\tau^e(\lambda)$, $\rho^e(\lambda)$ and $\alpha^e(\lambda)$, Fig. S1b. (In calculations, $l$ was equal to 3.8 mm.)

In order to simulate our approximate retrieval procedure, we will further multiply the spectrum $\tau^e(\lambda)$ by a certain scaling factor $x$ to make the corrected transmittance $\tau^c$ to be equal to unity in the spectral ranges that do not contain absorption lines, Fig. S1b. (This procedure is possible in rare-earth doped crystals characterized by narrow absorption lines and spectral ranges with practically no absorption, $k_{abs}l$<<1, between the groups of lines.) This scaled transmission $\tau^c$ can be viewed as the transmission measured between the two planes situated just inside the front and the rear walls of the sample. The common (although not precisely correct) practice is to retrieve the spectrum of absorption coefficients $k_{abs}(\lambda)$ using the formula

$$k_{abs}(\lambda) = -\ln(\tau^c(\lambda)/l), \qquad (S13)$$

where $l$ is the length of the crystal. As one can see in Fig. S1a, the retrieved *approximate* spectrum $k_{abs}(\lambda)$ is almost indistinguishable from the original *exact* spectrum $k_{abs}^e(\lambda)$. The retrieved spectrum $k_{abs}(\lambda)$ can be substituted to Eqs. 3, 4 and 1c, to calculate the *approximate* spectra $\rho(\lambda)$, $\tau(\lambda)$ and $\alpha(\lambda)$, which are also almost the same as the original *exact* spectra $\rho^e(\lambda)$, $\tau^e(\lambda)$ and $\alpha^e(\lambda)$, Fig. S1b. This good agreement proves high accuracy of the approximate retrieval method described above. We use this technique in our studies to retrieve the spectra $k_{abs}(\lambda)$ and $\alpha(\lambda)$ from the experimentally measured normal incidence transmittance spectra of the Er doped crystals (see Sections 2.2 and 6.3).

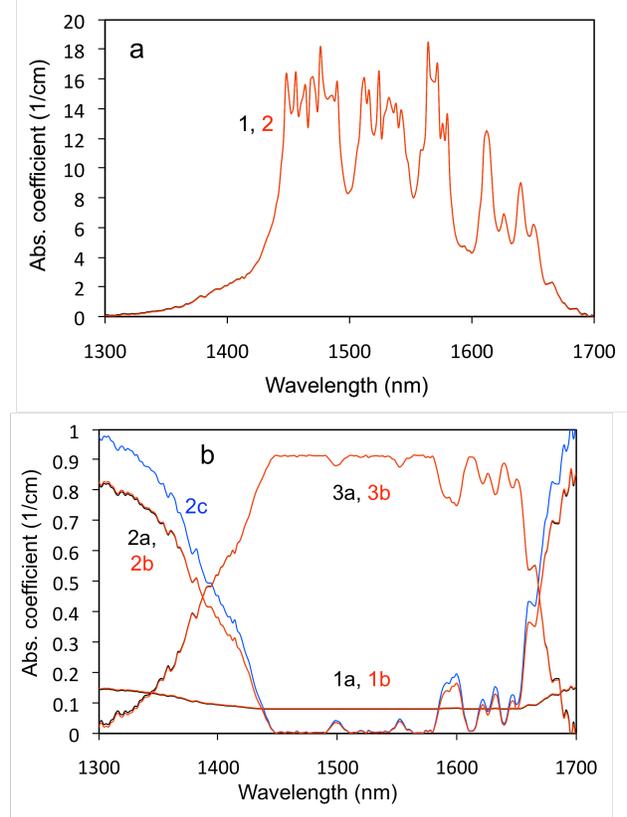

Fig. S1. (a) The original spectrum of absorption coefficients $k_{abs}^e(\lambda)$ (trace 1) and that retrieved using the approximate method (Eq. S13). The two spectra are almost on the top of each other. (b) The exact spectra $\rho^e(\lambda)$, $\tau^e(\lambda)$ and $\alpha^e(\lambda)$ (traces 1a, 2a and 3a, respectively) calculated using the spectrum of absorption coefficients $k_{abs}^e(\lambda)$. The analogous spectra $\rho(\lambda)$, $\tau(\lambda)$ and $\alpha(\lambda)$ (traces 1b, 2b and 3b, respectively) calculated using the spectrum of absorption coefficients $k_{abs}(\lambda)$. The scaled transmittance spectrum $\tau^c(\lambda)$, calculated as described above Eq. S13 (trace 2c).